# X-Driven Methodologies for SOA System Development – A Survey

Agustinus Andriyanto[1], Robin Doss[2], and Suhardi[3]

[1]*West Java Provincial Government, Indonesia, Regional Secretariat*
[2]*Deakin University, Geelong, Australia, School of Information Technology*
[3]*Bandung Institute of Technology, Bandung, Indonesia, School of Electrical Engineering and Informatics*
Corresponding author, e-mail: a.andriyanto@research.deakin.edu.au[1]

**Abstract**

**Purpose** – This study aims to evaluate four service-oriented architecture (SOA) system software development methodologies: business-driven development, model-driven development, event-driven development, and domain-driven development. These methods, generically labelled as x-driven methodologies (XDMs), are commonly used in a general software development context, but software architects can also apply them in an SOA-based system. Each XDM typically focus on a specific aspect that drives its processes and steps. This aspect is indicated by its label.

**Design/methodology/approach** –An evaluation method called qualitative screening mode is used in this study. XDMs are analysed based on their features to determine the suitability or support for service-oriented solutions. Criteria used to appraise each method are taken from SOA characteristics and SOA manifesto points.

**Findings** – Of the four discussed XDMs, business-driven development is the best-suited approach to implement a service-oriented system shown by its conformity with the selected assessment criteria. Nevertheless, the other three XDMs have also their own strengths. Model-driven development is excellent for productivity, event-driven development is preferential for a quick response and asynchronous work, while domain-driven development is distinctive to describe problems precisely.

**Originality/value** – The originality of this research is in the assessment general software development approaches of XDMs to be applied to SOA approach. The results can help developers in considering suitable methods to construct a prospective software system. Previous studies only investigate on methodologies designed intentionally for service-oriented systems.

**Paper type** – Literature review.

## 1. Introduction

Service-Oriented Architecture (SOA) is regarded as the next stage of distributed computing after several previously introduced concepts or methodologies (e.g., manual communication among the processes on different machines via tape in 1960s and '70s, followed by real-time access via low-level socket, and then by remote procedure calls and distributed objects in1990s and 2000s). The emergence of SOA was expected to solve the problems and restrictions related to RPC-based solutions like tight coupling between local and remote systems and the requirement of highly predictable networks due to their fine-grained nature and incompatible data types among programming languages (Davis 2009).

Several service-oriented software engineering (SOSE) methodologies are discussed and compared in the literature (Ramollari et al. 2007), (Kontogogos and Avgeriou 2009), (Gu and Lago 2011), (Svanidzaitė 2012), (Emadi et al. 2012). However, the previous studies are limited to the referred works that focus only on methods deliberately intended to SOA development. The studies affirm that the reviewed SOA processes are not revolutionary due to the adoption of commonly known methods, techniques and notation. Among those adopted methods are Extreme Programming (XP) and Rational Unified Process (RUP). Nevertheless, the writings do not address further the established software development methods which are adopted in SOA development processes. There are also not many works that compare and discuss them in relation to SOA. These include several development approaches focusing on one central aspect on which all the activities of software development are based and driven by the aspect. This study regards those methods as a group namely "x-driven" development methods, and it surveys four of them: business-driven development (BDD), model-driven development (MDD), event-driven development (EDD) and domain-driven development (DDD). In fact, the four x-driven methodologies (XDMs) are general software engineering approaches that are not intended purely for SOA development but are applied to many SOA projects like in (Koehler et al. 2008, Hentrich and Zdun 2016, Marzullo et al. 2008b, García et al. 2010, Lan et al. 2015, Levina and Stantchev 2009a, Autili et al. 2013, de Castro et al. 2012). Therefore, instead of adopting the criteria offered by previous papers discussing SOA methodologies, this study selects prominent SOA characteristics as well as the points declared in the SOA manifesto and explores how each quality is supported by each x-driven approach.

The paper is organised as follows. Section 2 introduces SOA concepts, software development methods and survey papers relating to SOA development. Then, the next part presents how the XDMs are selected for this study along with the evaluation method and its criteria. The four XDMs are subsequently described in detail in Section 4. Section 5 examines these methodologies to study how they align with SOA characteristics and SOA manifesto. Finally, this study concludes the work in section 6.

## 2. The Survey Papers of SOA Development Methodologies

*2.1. SOA Development Methodologies*

(Erl 2005) defines SOA as the representation of automation logic comprising smaller, distinct units of logic where these units are distributed individually but collectively cover a larger piece of business automation logic. SOA is a design for realising the vision of Service Oriented Computing (SOC) paradigm (Papazoglou et al. 2007). It is a more progressive construct than the traditional software-as-a-service (SaaS) concept in which a company outsources its information technology requirements through the internet on a monolithic and tightly coupled solution. Although the services in SOA can be implemented by different platforms, mostly they are realised by web services. Two mainstream technologies that form the basis of web service construction are SOAP and REST. A study by (Belqasmi et al. 2012) shows that the newer technique RESTful style services offer better performance than the SOAP-based services. REST simplifies the request and message processing method and uses the only HTTP as the application-layer protocol. In more recent years, a concept called microservices architecture (MSA) (Newman 2015, Dragoni et al. 2017) is introduced to tackle the complexity of SOA. While services are designed to be loosely-coupled in the traditional SOA, MSA focus only on a single functionality implemented as an independent service. Due to its similarity in characteristic with the traditional SOA, MSA is often regarded as a specific architectural form of SOA.

Based on the conducted software development processes, e.g. planning, modelling, construction, and deployment, many software process models have been proposed known under software development life cycle (SDLC) models. Some models gain enough popularity such as waterfall model, unified process model, and agile model. Several methodologies are named based on a specific aspect which drives the whole process, e.g. model-driven development, business process-driven development, data-driven development. These methodologies still embrace one of SDLC models for the step-by-step processes performed. Other methodologies focus on the building block elements of a system to produce "element-oriented" development methodologies such as object-oriented development or service-oriented development (SOD).

This study refers to SOD for any methods producing an application based on service-oriented architecture concepts. The development focuses on business processes embodied in reusable building blocks (i.e. services) that are self-describing and independent of computing platforms (Papazoglou and Van Den Heuvel 2006). In SOD the services are defined at a higher level of abstraction, and they can be realised into a procedure-oriented language, a message queuing system, or an object-oriented system through service definition mapping (Newcomer and Lomow 2005). As the fundamental elements for developing applications/solutions, services provide well-defined interfaces or service descriptions for interacting with other services or potential consumers (Papazoglou 2003). The logic encapsulated in a service can be a business task, a business entity or some other logical grouping that vary regarding size and scope (Erl 2005). Based upon the service description, services communicate via messages in loosely coupled relationship. The basic SOA involves three kinds of participants, i.e. service provider (SP), service registry (SR), and service client (SC) (Papazoglou 2003). Then, the three basic operations are related to the interaction between them: publish for SP-SR interaction, discover or find for SR-SC interaction, and invoke or bind for SP-SC interaction.

The position of SOA development method in relation to the other methods is described by (Koskela et al. 2007), contrasting SOA with three different software development paradigms, i.e. Component-Based Development (CBD), Object-Oriented Programming (OOP) and Aspect-Oriented Programming (AOP). Each of the approaches takes a different position regarding the layered organisation of artifacts or elements. Owning the smallest artifacts, OOP is at the lowest layer, followed by CBD locating at the middle of the organization. SOA has the largest artifacts scope which lies at the top. AOP is positioned hand in hand with the others on all the layers since it concentrates on certain recurrent aspects, not reusable building blocks.

Several comprehensive methods are proposed with step-by-step processes like in software development life cycle methods, e.g. Thomas Erl's method (Erl 2005), Papazoglou method (Papazoglou and Van Den Heuvel 2006), IBM SOMA (Arsanjani 2004), and SOUP (Mittal 2005). A method adapts a general software development methods and considers the fundamentals of agile software development and assess the approach suitability for SOA-based systems (Krogdahl et al. 2005) or presents five steps to add agility to SOA methodologies (Shahrbanoo et al. 2012). Meanwhile, the others adapt one of the 'x-driven' methodologies, so the names of the approaches correspond to the method engaged, e.g. event-driven SOA development, model-driven SOA development or policy-driven SOA development.

*2.2. The Survey Papers*

Papers introducing SOA was started in 2002. Several years later, service-oriented software engineering methods were introduced, followed by some survey papers that compared and evaluated those methods. A survey paper by (Ramollari et al. 2007) uses a relative quantitative scale from 1 to 5 for some of the criteria to measure the degree of offered aspects such as prescriptiveness and agility. It is concluded that IBM SOMA, CBDI-SAE and Thomas Erl's methodologies are the most prescriptive approaches while SOUP is the most agile. Another study by (Gu and Lago 2011) compares some approaches with two criteria groups: general and service-specific aspects. While other investigations cover the complete lifecycles of software development, (Svanidzaitė 2012) makes an evaluation that only focuses on analysis and design phases. There is also a study from (Emadi et al. 2012) that not only assesses the existing approaches but also offers an improved SOA methodology over the approaches. See table 1 for detailed number and name of the methods evaluated in each study.

Aside from the survey papers evaluating the SOA development methodologies, there are also some works that contrast different viewpoints of software development to build an SOA system or distinguish the approaches from other development paradigms. A comparative study by (Al-Rawahi and Baghdadi 2005) presents two perspectives on developing web services, i.e. IT-oriented and business-oriented perspectives. Another work compares SOUP as one of SOA development methodologies with two established software design and development methodologies (SDDM), i.e. RUP and XP (Svanidzaitė 2014).

**Table 1.** The survey papers discussing and comparing SOA development methodologies

| Author | Number of Analysed SOA Methodologies | Name of SOA Methodologies |
|---|---|---|
| (Ramollari et al. 2007) | 10 | SOAD, SOMA, RQ, SAE, SOAF, SOUP, Pap., Erl, B2B, SJ. |
| (Kontogogos and Avgeriou 2009) | 7 | SeC, SOD, Pap., SOAF, SOMA, True, SAE. |
| (Gu and Lago 2011) | 12 | SAE, Pap., SeC, SOMA, SJ, SOAD, SOSE, SOUP, True, Cha., SOD, SOAF. |
| (Svanidzaitė 2012) | 5 | SOMA, SOAF, Pap., Erl, SOUP. |
| (Emadi et al. 2012) | 9 | SOAD, SOMA, RQ, SOAF, SOUP, Pap., SJ, RUP, MSM. |

Note:
SOAD = IBM Service Oriented Analysis and Design (Zimmermann et al. 2004)
SOMA = IBM Service Oriented Modelling and Architecture (Arsanjani 2004)
RQ = SOA RQ (Repeatable Quality) by Sun Microsystems (Microsystems)
SAE = CBDI-SAE (Service Architecture and Engineering) Process (Allen 2007)
SOAF = Service Oriented Architecture Framework (Erradi et al. 2006)
SOUP = Service Oriented Unified Process (Mittal 2005)
Pap. = Papazoglou's Methodology (Papazoglou and Van Den Heuvel 2006) a.k.a SDLM (Service Development Lifecycle Methodology)
Erl = Thomas Erl's Methodology (Erl 2005)
B2B = BPMN to BPEL (Emig et al. 2006)
SJ = Steve Jones' Service Architectures from OASIS (Jones and Morris 2005)
SeC = Service-centric System Engineering (SeCSe) (SeCSe-Team 2005)
SOD = Service-Oriented Development In a Unified fraMework (SODIUM) (Topouzidou 2007)
True = True SOA from G. Engels et al. (Engels et al. 2008)
SOSE = Service-Oriented Software Engineering Framework (Karhunen et al. 2005)
Cha. = Chang's Methodology (Chang 2007)
RUP = RUP for SOA (Hussain et al. 2010)
MSM = Mainstream SOA Methodology that refers to Erl's Methodology.

## 3. Research Methodology

*Selecting X-Driven Approaches*

Although there are many XDMs for general software development, not all of them are frequently mentioned in published papers. This study found that only four of the approaches are quite popular and they will be the focus of the discussion: business-driven development, model-driven development, event-driven development and domain-driven development. Other XDMs might be involved in some SOA developments as well, such as policy-driven development (Demchak and Krüger 2012), metadata-driven development (Pintar et al. 2009), and capability-driven development (España et al. 2014). However, since they do not get much attention in the literature, this paper does not include them in the discussion and set the context for only the most popular ones.

*Evaluation Method*

This study uses an evaluation method called qualitative screening mode to assess the four x-driven SOA development methodologies. The method is one of the nine evaluation methods defined by (Kitchenham et al. 1997) to evaluate software engineering methods and tools. Being the results of the DESMET project, those methods are categorised into three evaluation types commonly applied in many scientific fields, i.e. quantitative, qualitative and hybrid. Methods belonging to the last type have some quantitative and qualitative elements, e.g. benchmarking in which the measured performances can be quite objective, but the determination of the selected specific tests is subjective. Another DESMET methods categorisation is based on the three evaluation procedures. The first is *formal experiment* where users try the approaches by doing some tasks before making their evaluations. *Case study* as the second category is for evaluating approaches applied on a real project, while in the last one, i.e. *survey*, users give information/opinion after they have used or studied the approaches. The nine DESMET evaluation methods are combinations of distinct types and procedures. Two kinds of them under *qualitative type* with *survey procedure* are *qualitative survey* and *qualitative screening*. Although both are similar and suitable for a feature-based evaluation, they are different concerning the number of people doing the work. While the *qualitative survey* is done by a group of people who have used or studied the examined approaches, the *qualitative screening* is performed by a single individual only. That person not only undertakes the assessment but also sets the features to be assessed and their rating scale. This study is a literature review carried out by the authors and based on written works describing the approaches rather than the actual use of the approaches. Therefore, the authors consider that *qualitative screening* is the most suitable method for evaluating the characteristics of the observed X-driven methodologies.

*Assessment Criteria*

To evaluate software development methodologies, assessment criteria are needed, and there are papers suggesting the criteria either for general software or SOA development. Some papers present their criteria as a supporting part of analysis such as (Al-Rawahi and Baghdadi 2005), (Svanidzaitė 2014), or SOA-evaluation papers listed in table 1. Some others deliberately propose the criteria that might become references for other papers like in (Gholami et al. 2010) or (Baghdadi 2012). As this study does not review SOA-specific development approaches or SDLC models, the evaluation criteria are selected based on the general characteristics of the x-driven methods. The assessment involves the SOA main characteristics as described in (Erl et al. 2017) and the SOA manifesto points (Arsanjani et al. 2009) to examine the XDMs conformity with the service-oriented construct.

*Methodology Limitations*

As with other evaluation methods, the qualitative screening approach also comes with some inherent limitations. First, the evaluation criteria are subjective. There may be other criteria which are not included in the analysis but contribute to the qualities of the observed subjects. Another limitation is related to the information on which the assessment is based that comes from third parties. Furthermore, this method does not consider XDMs actual implementation experiences or case studies that are more proven (although the case study method has some limitations as well). It only focuses on how the reviewed approaches achieve specific criteria based on the subjective evaluation of the authors. The finding consequently depends on the selected criteria, secondary source information, and the authors' view. However, this study tries to alleviate this problem by setting the criteria from general features that are commonly mentioned in the literature and considered to be mainly accepted. The authors also attempt to refer to any significant works concerning the topic under study and use those sources to justify the finding.

## 4. X-Driven Development Methodologies

### 4.1. Business-Driven Development (BDD)

*BDD Term and Objectives*

The term "Business-Driven Development" (BDD) is used interchangeably with business-process-driven development or process-driven development. It focuses on how business processes align with IT with the primary objective to automate and support business processes. Introduced by (Mitra 2005) in 2005, business-driven development defined as a methodology for developing IT solutions that aim directly to satisfy business requirements and needs. This approach centres on content or business processes, not the technology with BPM and BPEL as the popular tools. Understanding business requirements, creating BPMs through the as-is and to-be models, and employing the BPMs in the design and development phases are the key concepts.

*BDD and SOA*

The SOA Consortium managed by OMG outlined the meaning of "business-driven SOA" in their released whitepaper (SOA-Consortium 2010). The document explains the two-fold or mutual relationship between business architecture and information technology: business architecture defined as the formal representation and active management of business design is a critical input to IT while IT trends and capabilities influence business design choices. This notion leads to what is called as a true enterprise architecture practice that gives equal emphasis to business and technology concerns. How BDD meet with SOA is discussed in (Juric and Pant 2008) and (Hentrich and Zdun 2011). (Hentrich and Zdun 2011) defines the term "Process-driven SOA" as an architecture in which services that consist of service-based middleware and process engines are called to implement business processes. Generally, one basic service in SOA represents one business function, and one composed service represents several collaborating business functions constituting a business process. Therefore, BDD that observes a system more from a business viewpoint promotes one of the service-oriented application goals, i.e. to abstract the business logic away from its low-level system implementation. Some issues regarding BDD and SOA cover development lifecycles (Mitra 2005, Juric and Pant 2008), business services modelling and analysis (An and Jeng 2007), and model transformation (Koehler et al. 2008).

*4.2. Model-Driven Development (MDD)*

*MDD Term and Objectives*

The term "Model-Driven Development" is used interchangeably with Model Driven Architecture (MDA), Model-Driven Software Development (MDSD) and Model Driven Engineering (MDE) and they frequently refer to the same thing. However, they also have a slightly different scope, constraints, and meaning (Sommerville 2011, Stahl et al. 2006). According to Mellor et al. (Mellor et al. 2003), MDD is "'the notion that we can construct a model of a system that we can then transform into the real thing'". (Selic 2003) defines MDD as a software development methodology focusing on models as the primary products rather than computer programs. MDD is known as a set of software development approaches based on models, modelling, and model transformation (Brown et al. 2005). Two fundamental ideas behind the MDD are abstraction and reuse (Fabra et al. 2012). Abstraction allows the system designed on different levels of representation in a set of models while reuse enables software components implemented in diverse environments since the models are separated from specific implementation technologies or platforms.

*MDA by OMG*

MDA (Soley 2000, Miller and Mukerji 2001, Miller and Mukerji 2003) is a standard of MDD methodology defined by the Object Management Group (OMG). It defines three models for abstracting a system into different viewpoint levels:
1. Computation Independent Model (CIM)
   Sometimes called as a domain model or business model, the CIM does not represent the system structure or process but focuses on the system requirements and environment.
2. Platform Independent Model (PIM)
   As opposed to CIM, the PIM defines a complete specification of the system structure and process but hides the implementation details of a particular technology or platform.
3. Platform Specific Model (PSM)
   The PSM augments the complete specification in the PIM with the details that define the system implementation on a particular technology or platform.

*MDD and SOA*

Seeing that the main output of MDD is a model, some reference models in which services act as the first class are proposed, e.g. SoaML, SOA-RM, and SOA Ontology. (Mohammadi and Mukhtar 2013) provides a review and comparison of 7 SOA modelling approaches for enterprise information systems. The topic of MDD for SOA development is very popular that more than one hundred papers are published as listed in (Ameller et al. 2015) conducting a mapping study of related papers from 2003 to 2013. This research examines the papers included in (Ameller et al. 2015) and also other recent papers related to the topic and organise them according to the main ideas discussed in each paper. The papers are then categorised into four large topic groups as illustrated in table 2.

**Table 2.** Topic Groups of MDD in SOA and The Number of Published Papers from 2003-2016

| | | |
|---|---|---|
| MDD-SOA (139) | General implementation (35) | General system (25) |
| | | General system with model only (3) |
| | | Specific system (5) |
| | | General concepts without model and transformation (2) |
| | Service composition (20) | Service composition in general (13) |
| | | Service orchestration (4) |
| | | Service choreography (3) |
| | Incorporating MDD-SOA with other concepts/issues (45) | Model and transformation (13) |
| | | Context-aware services (7) |
| | | Migration and integration (8) |
| | | Other concepts/issues (17) |
| | Non-functional properties (NFPs) (39) | General NFPs (9) |
| | | Security (15) |
| | | Other NFPs (15) |

Some of the many proposed MDD-SOA frameworks are SOD-M (De Castro et al. 2009, De Castro et al. 2011) and ODSOMDA (Zhang et al. 2012). IBM also provides a model-driven design of SOA solutions called SOMA-ME (Service-Oriented Modelling and Architecture Modelling Environment) (Zhang et al. 2008). Other methodologies in MDD-SOA are proposed as well such as CAIDE (Bahler et al. 2003), STRIDE (Bahler et al. 2007), ArchiMeDes (López-Sanz and Marcos 2012), and OOWS (Ruiz et al. 2005), (Ruiz et al. 2006), (Ruiz and Pelechano 2007).

*4.3. Event-Driven Development (EDD)*

*EDD Term and Objectives*

EDD is linked with a concept of EDA (Event-Driven Architecture) as a complementary architecture of SOA (Schulte and Natis 2003). As service-oriented development is a way of producing service-oriented applications, event-driven development is a methodology to build IT solutions based on an event-driven architecture. EDA is defined as "structure in which elements are triggered by events" while an event is defined as "a change in the state of one of the business process elements, which influences the process outcome" (Levina and Stantchev 2009b). IBM defines the meaning of EDA as an approach in which applications and systems are designed and implemented to enable event delivery between decoupled software components and services (Maréchaux 2006). It can be viewed as a system architecture that manages and executes "when...then..." rules, i.e. when reality deviates from expectations or an important situation (event) occurs, then update expectations and respond appropriately (Chandy 2006).

*EDD and SOA*

Event-driven Service-oriented Architecture (EDSOA) can be viewed as the combination of event-driven architecture (EDA) and SOA. It is defined as a type of SOA in which events are responded to by services that can be triggered in the beginning or end of service execution, changes in critical variables value or exceptions (Laliwala and Chaudhary 2008). EDA and SOA collaboration can be summarised in three points (Malekzadeh 2010): (1) Shared use of the data source and data semantics, (2) EDA uses SOA services to manage the information, (3) SOA uses EDA to distribute and publish information. Some examples of EDSOA implementations are in IoT (Zhang et al. 2014, Lan et al. 2015), space command and control (Yuan and Watton 2012), and enterprise systems (Zagarese et al. 2013).

EDSOA can also be integrated with MDA (model driven architecture) to form MDA-SOA-EDA (Sriraman and Radhakrishnan 2005). In the future, large-scale distributed applications will involve more perceived services (the services in which the execution is driven by events), so EDSOA is regarded as the next generation of SOA (Lan et al. 2015). This approach was also named as SOA 2.0 by Oracle. However, not all systems are suited for event-driven architectures especially when the nature of the processes is based mainly on a request/response relationship.

*4.4. Domain-Driven Development (DDD)*

*DDD Term and Objectives*

Domain-driven design (DDD) is a software development methodology for complex needs by focusing on the evolving domain model, i.e. the model of automated business processes or real-world problems. It is mainly about people and communication instead of technical issues and giving more attention to core business rather than a software application. The DDD was based on the assumption that "the heart of software development is knowledge of the subject matter" (Evans 2004).

As the main part of this approach, a domain model is a ubiquitous language that structures the problem and speaks both the user and the developer languages. The language can be an ad hoc diagram, UML, document or any other forms as long as it can be understood by domain experts, software architects and software developers. The concept of service in DDD has similarities with SOA as a stateless operation stands alone in the model with its defined interface. However, the granularity is "medium-grained" since it is built in the context of a software system instead of "coarse-grained" SOA that is commonly implemented in different organisations. DDD helps the involved stakeholders to diminish both the domain gap (between domain expert and designer) and the model gap (between designer and programmer). The proposal from (Le et al. 2016) is one of the methods to build a ubiquitous language employing an object-oriented design that uses meta-attributes to build the domain class model.

*DDD and SOA*

Perspectives differ on DDD and SOA providing varying degrees of relationship between them. The major viewpoint considers DDD and SOA as two concepts with different scope targets, but they are highly complementary. While the coverage of DDD is at application scope, SOA manages and integrates multiple deployment units at an enterprise or inter-enterprise scale. A coarse-grained service in SOA, therefore, represents a well-design unit which encapsulates the domain models built using DDD approach.

Even though DDD is quite famous as one of software development methodologies, few papers are discussing this topic as an approach to build SOA-based system. One of them, (Marzullo et al. 2008a) implemented this methodology for SOA development combined with MDA for organisations sharing the common business field. DDD is one of the pillars in which microservice architectures (MSA) are built in addition to SOA and object-oriented programming. It has a similarity with one of the MSA objectives, i.e. to manage complexity by identifying core and auxiliary domains. Managing complexity is in accordance with SOA intention as well, i.e. to solve the challenges of the large monolithic applications.

## 5. Discussion

*5.1. Characteristics of The X-Driven Methodologies*

The focus or approach of each methodology can be observed in most cases from their respective names. Table 3 summarises the characteristics of each methodology.

*BDD*

Business-driven development is proposed as an improvement of the previous point of view on enterprise architecture that accentuates on technology, and it attempts to balance the method with business architecture (SOA-Consortium 2010). Some benefits gained from BDD are the reduced gap between business processes and IT, fewer system faults and faster development phases that reduce delivery times for fulfilling business demands (Juric 2010). However, since it focuses more on a higher level, i.e. business processes, a software architect putting into practice this methodology must be aware of the details on the system level, keeping in mind BDD's objective to manage flexible businesses through flexible IT solutions.

**Table 3.** General characteristics of BDD, MDD, EDD and DDD

| Characteristic | Business DD | Model DD | Event DD | Domain DD |
|---|---|---|---|---|
| Purpose and objectives | Alignment between business and IT | Automatic transformation from models into program code | Quick adjustment to unpredictable system and environment changes | Precise problem description using ubiquitous language |
| Primary focus | Content or business process, not technology | Model as principal output | Events triggering the system execution | Core domain and creating a domain model |
| Suitable environment | System with large number of different types of users with different purposes of use | Several similar systems in different environments for increasing productivity | System with asynchronous work, information flow and mission-critical | System with complex requirements when a conceptual model is iteratively refined |
| Defined for the first time by | Mitra (2005) and OMG (2005), then by SOA-Consortium (2010) | OMG (2001) | Schulte & Natis (2003) | Eric Evans (2003) |
| Lifecycle coverage | Full (5 phases: model, develop, deploy, monitor, analyse & adapt) (Mitra 2005) | Design & implementation | Full (4 phases: model, compose, deploy, execute & monitor) (Laliwala & Chaudhary 2008) | Full (5 phases: model, design, develop, testing, refine-and-refactor) (Penchikala 2008) |
| Main Artifacts/products | Business requirements | CIM, PIM, PSM, model-mapping | Event flow layers (event generator, event channel, event processing engine, downstream event-driven activity) | Domain model, i.e. ubiquitous language |
| Common Notation/tools | BPMN and BPEL | UML | UML | UML |
| Delivery strategy | Top-down (to-be) and bottom-up (as-is) | Mostly top-down but can be bottom-up (reverse-engineering) | Outside-in | Top-down |
| Problem/solution focus? | Problem and solution balance | Solution | Solution | Problem |

*MDD*

Increasing productivity is the primary benefit of model-driven development as it enables developers to generate models and code automatically. Additionally, a system can be understood more easily as models create several abstraction levels of the system and represent up-to-date domain knowledge and documentation. The cost for modelling and transformation, however, must be considered, particularly if the system built is only a single solution implemented on a single platform. Redundant or duplicate artefacts might cause inefficiency both in development and maintenance time. Therefore, this method is better implemented on several systems having similarities or a system evolving frequently. Incorporating MDD into other x-driven approaches is common.

*EDD*

An event-driven architecture has an inherently extreme loose coupling, quick response and highly distributed services. The ideal cases for this approach are therefore concurrency execution, asynchronous work and information flow as well as real-time and mission-critical applications like IoT (Internet of Things) or context-aware system that provide services based on user actions (Maréchaux 2006, Michelson 2006, Sriraman and Radhakrishnan 2005, Zhang et al. 2014). However, since events are mostly independent while business activities are rarely so, events do not provide powerful expressiveness to describe business logic.

*DDD*

Because of its nature, the domain-driven development approach can only be implemented well in longer term and complex projects. Knowledge of business process is essential more so than technical skills, and the development process is performed iteratively and refined continuously potentially leading to an expensive project. As business experts and developers in DDD shared their knowledge of domain through models on which MDD focuses, there is a large intersection between DDD and MDD. Therefore, the domain model is represented in MDD by the structure and design of a system. As stated by (Klein 2007), the domain model and the ubiquitous language supporting it are the main focus of DDD while projects using MDD start with Domain Specific Language (DSL) and a model facilitating the DSL.

*Summing up The XDMs Characteristics*

Aside from their different characteristics, the XDMs have commonalities as well. BDD and DDD have a similarity in that the focus lies on business aspects, and therefore these methodologies deal with a higher abstraction layer of the system. Their intentions, however, are distinct as BDD aims to bridge the gap between business and IT while DDD's objective is to depict the real problem using a ubiquitous language which can be understood by both domain experts and developers. It is possible to apply some concepts in one approach to another approach such as DDD that can be employed to answer one of the BDD's main challenges, i.e. making a business-process model that precisely describes the required IT services. Some practices make use of those x-driven methodologies simultaneously such as (Hermawan and Sarno 2012) who employs BDD, MDD and DDD to develop an SOA-based system combined with resource-oriented architecture (ROA). While BDD, MDD and DDD can be a solution for a software system in general, EDD has specific characteristics which fit on a proactive, asynchronous, event-triggered-based environment.

*5.2. Suitability for SOA Development*

*5.2.1. Suitability with SOA characteristics*

Compared to other XDMs, model-driven development is the most widely discussed and used in SOA development that more than one hundred papers covering this topic have been published with many supporting tools and standards. On the contrary, not many papers discuss the other three methods. The suitability of each method is examined with some SOA properties as defined by (Erl et al. 2017) depicted in table 4. Only four major characteristics are mentioned in this case despite many characteristics listed in many works of literature, i.e. business-driven, vendor-neutral, enterprise-centric, and composition-centric.

*Business-driven*

Business-driven as the first key attribute means that a solution built is not only for fulfilling tactical or short-term business requirements but also for anticipating strategic or long-term business goals. BDD support this feature through as-is and to-be models. Since the alignment between business and IT was an initial SOA promise, the business process-driven development method is highly suited to the characteristic as this approach tries to bridge the gap between those two entities. Changes in business needs are accommodated through the analysis-and-adaptation step in the BDD execution model. Likewise, MDD accommodate business requirement changes due to explicit links between the requirements and its models for which the code and other lower-level models can be adjusted automatically. The business aspect is also a major concern in DDD in which business domain concepts are mapped into software artifacts such as entity, value object, aggregate etc. That model will be iteratively refined over time to keep pace with changes in business requirement as well as the corrected understanding of problem domain. On the other hand, the nature of EDD requires much effort to adjust the evolving business needs both on client and server (Schulte and Natis 2003).

*Vendor-neutral*

MDD is in accordance with one of SOA objectives to be technology agnostic. With its levels of abstraction through CIM and PIM, MDD allows a system to be defined regardless of their implementation platforms. BDD and DDD however also adopt MDD as part of their methodology, so they promote the second characteristic as well although it is not their main goals. DDD has a similar characteristic with BDD in that it works on higher-level of a system as well as EDD which based on an abstract software structure namely EDA.

*Enterprise-centric*

In the third characteristic, i.e. enterprise-centric, the logic of solutions built are targeted as reusable enterprise resources and not attached to a specific implementation boundary. EDD that has the distributed architecture as its nature supports this enterprise-centric feature even though it does not guarantee the avoidance of new silos within an enterprise. BDD accommodates the characteristic as well since one of its key aspects is the identification of reusable assets. However, DDD to some degree must maintain duplicate resources because of its goal of reducing complexity in the software.

*Composition-centric*

Composition-centricity as the last feature needs flexible resources that can be incorporated into various aggregates structures but it seems that all four methodologies do not strongly support this feature or explicitly state it as one of their attributes.

It should be noted that EDD has a unique position among other methodologies as EDA has distinct differences to SOA. While SOA features loosely coupled interactions, one-to-one communications, consumer-based trigger and synchronous response mode, EDA accentuates completely decoupled interactions, many-to-many communications, event-based trigger, and asynchronous response mode (Lan et al. 2015). SOA and EDA, however, have similarities as well, i.e. (Schulte and Natis 2003) modularity with reusable business components, designed for distributed systems and connections for program-to-program communication, and implemented mainly through web services.

*5.2.2. Suitability with SOA Manifesto*

How the x-driven development methodologies align with the objectives of SOA is also investigated using the points in SOA manifesto (Arsanjani et al. 2009) shown in table 4. The first two of the manifesto could be regarded as similar, and hence both items are strongly supported by BDD while DDD moderately promotes them. Although MDD is adaptive to business changes, it does not support these points strongly as the main concern is producing software models which is more related to technical matters.

The third value of the manifesto requires that software programs be natively compatible and they do not need to be customised for integration as it can be costly, time-consuming and lead to unwanted complexity. In the service concept, this is fulfilled by separating service interface with its implementation. None of the methodologies directly emphasis on interoperability but MDD has many standards for service modelling that can be actualised into inherently compatible applications. The next priority value requires multi-purpose logic that can be shared and reused across business processes, applications and enterprises. This quality resembles enterprise-centric feature sustained by most of the discussed methods. However, the service reuse principle in SOA has a potential drawback to DDD because it can make some elements as the manifestation of domain models become out of context.

Flexibility as the next point enables a solution to accommodate ever-changing business requirements with minimal effort. All the discussed methods strongly support this feature. BDD with its development lifecycle in which the last step (analyse and adapt) allows user-needs alteration. MDD with the multilevel models enables modification in one model will be propagated automatically down to the code level. Meanwhile, the nature of EDD provides ease for reassembling and reconfiguring. DDD offers flexibility too as it is heavily based on the concepts of object-oriented analysis and design in which business logic is encapsulated in aggregates, making a system easier to be altered. The last point of the manifesto is quite similar with the previous one, i.e. the solution built should be flexible to adapt to business processes that change continuously, so all the approaches provide the same support to this feature.

**Table 4.** X-Driven Development Methodologies support to SOA

| Criteria | Business DD | Model DD | Event DD | Domain DD |
|---|---|---|---|---|
| Suitability with SOA objectives/natures | Alignment between business and IT | Not concern with technological details | Designed for distributed systems and program-to-program communication | Managing complexity to solve the challenges in a large monolithic application |
| Support SOA Characteristics: | | | | |
| • Business-driven | ☑ | ☑ | ☐ | ☑ |
| • Vendor-neutral | ☑ | ☑ | ☑ | ☑ |
| • Enterprise-centric | ☑ | ☑ | ☑ | ☐ |
| • Composition-centric | ☐ | ☐ | ☐ | ☐ |
| Support SOA Manifesto: | | | | |
| • Business value over technical strategy | ☑ | ☐ | ☐ | ☑ |
| • Strategic goals over project-specific benefits | ☑ | ☐ | ☐ | ☑ |
| • Intrinsic interoperability over custom integration | ☐ | ☑ | ☐ | ☐ |
| • Shared services over specific-purpose implementation | ☑ | ☑ | ☑ | ☐ |
| • Flexibility over optimization | ☑ | ☑ | ☑ | ☑ |

| Criteria | Business DD | Model DD | Event DD | Domain DD |
|---|---|---|---|---|
| • Evolutionary refinement over pursuit of initial perfection | ☑ | ☑ | ☑ | ☑ |

## 6. Conclusion

Apart from existing approaches deliberately proposed for SOA development, there are also general software development methodologies that can be adapted to build a SOA system. Some of them focus on a specific aspect, acting as a central point for all activities in the development lifecycle, referred to as *x-driven development methodologies*. The four methodologies selected in this study show various supports to SOA system development. Each method might have a focus on one SOA feature without sufficiently addressing the other characteristics. Moreover, each method has its top priority that determines their main characteristic: business and IT alignment for BDD, productivity for MDD, quick response and asynchronous work handling for EDD, and precise problem description for DDD.

BDD seems to be is the most suited for SOA, as it is part of the proposed construct. Nevertheless, MDD is the most popular topic discussed in many papers, and it is common for other XDMs to employ the method. The qualities of EDD which are opposite to SOA characteristics give rise to its uniqueness compared to the others, but EDD is considered to play an important role in the future. Despite each method's advantages and disadvantages, a software architect can incorporate them to get a sum benefit obtained from each method while the drawbacks inflicted by one method are alleviated using another one.

This study sets the context to only four XDMs that are regarded as the most popular ones. There may be other XDMs that are suitable to be adopted in SOA development, but as they are not much presented in published papers, this paper does not include them in the discussion. Likewise, a new alternative method may require time to be recognised and incorporated in the SOA development. This paper intentionally only assesses the general characteristics of the methodologies and leave out the detail. The selected evaluation method, i.e. qualitative screening also has some limitations as it relies on the chosen criteria, secondary source information, and authors' viewpoint that may affect the finding. Nevertheless, this work has attempted to provide a coherent review of the existing x-driven methodologies engaged in SOA system development. Further studies will explore how the XDMs are combined with other approaches that could be either x-driven, general established software development, or one of SOA development methodologies.

**Acknowledgement**

Our sincere gratitude to LPDP (Indonesia Endowment Fund for Education) of Ministry of Finance for granting Doctoral scholarship to the first author to study at Deakin University. We would also like to thank West Java Provincial Government for other supports.